\documentclass{article}

\input pz.sty
\input epsf.sty
\usepackage{longtable}
\usepackage{graphicx}

\begin{document}

\begin{center}

\PZtitletl{PULSATIONS AND LONG-TERM LIGHT VARIABILITY} {OF THREE
CANDIDATES TO PROTOPLANETARY NEBULAE}

\PZauth{V. P. Arkhipova, N. P. Ikonnikova, G. V. Komissarova}
\PZinsto{Sternberg Astronomical Institute, University Ave. 13,
119992 Moscow, Russia}

\end{center}

\begin{abstract}

We present new photometric data and analysis of the long-duration
$UBV$ photoelectric observations for three candidates to
protoplanetary objects -- F-supergiants with IR-excesses located
at large galactic latitudes, IRAS $18095+2704$, IRAS $19386+0155$,
and IRAS $19500-1709$. All three stars have revealed quasiperiodic
low-amplitude variabilities caused by pulsations observed against
the long-term trends of brightnesses. For IRAS $18095+2704$=V887
Her we have found a pulsation period of 109 days and a linear
trend of brightness under the constant colours if being averaged
over the year timescale. The light curve of IRAS
$19386+0155$=V1648 Aql over 2000--2008 can be approximated by a
wave with a main period of 102 days which is modulated by close
frequency, with a period of 98 days, that results in brightness
oscillations with a variable amplitude. V1648 Aql has also shown
synchronous reddening together with a persistent rise of
brightness in the V-band. IRAS $19500-1709$=V5112 Sgr experiences
irregular pulsations with the periods of 39 and 47 days. The
long-term component of the variability of V5112 Sgr may be related
to the binary character of this star.

\end{abstract}

\section{Introduction}

After termination of the superwind stage at the tip of the
asymptotic-giant branch (AGB), a star with the initial mass of 1
to 8 solar masses remains with a carbon-oxygen (CO) core, of $0.5
- 0.8 M_{\odot}$, embedded into extended gaseous envelope which
imitates such properties of supergiant stars as luminosity and
gravitational acceleration. The star is also surrounded by the
dust envelope which becomes more transparent during the expansion.
The theoretical calculations by Bl\"{o}cker (1995) indicates the
following evolution of the star under the constant bolometric
luminosity: while the star contracts gradually, its temperature
rises. During the post-AGB stage of their evolution stars migrate
to the left at the Herzsprung-Russell diagram suffering
pulsational instabilities under certain temperatures (Gautschi
1993). Moreover, mass loss due to stellar wind persists at this
evolutionary stage.

The aim of our photometric observations of post-AGB stars is the
study of their variability origin: it may be pulsational activity,
long-term evolution, evolution of their dust envelope, variable
stellar wind, and/or binary status. In this work we present the
results of almost 20 years observing three supergiants with
IR-excesses -- the candidates to protoplanetary objects which may
represent the full illustration of various types of variability
among this class objects.

\section{Observations and Data Analysis}

Table 1 contains the global data on the stars under studying.
Besides the considerable IR-excess, their common properties are
also high galactic latitude, the spectral class F, and rather low
metallicity of their atmospheres.

Our photometric observations have been made in the Crimean
observatory of the Sternberg Astronomical Institute, at the 60 cm
Zeiss telescope, with photon-counting $UBV$-photometer. During our
observations we used the aperture of $27^{\prime \prime}$. The
accuracy of individual estimates was 0.01 to 0.02 magnitudes,
depending on weather conditions and star brightness. For two of
three stars, V1648 Aql and V5112 Sgr, we involve also the data of
the Automathic All-Sky Survey, ASAS (Pojmanski 2002). The analysis
of the periodicity of light variations has been made with the
software developed by V.M. Lyuty which is based on the
Fourier-analysis approach proposed by Deeming.

\begin{table}
 \begin{center}
\caption{List of objects}
\bigskip
\begin{tabular}{|c|c|c|c|c|c|c|c|}
\hline

IRAS&HD&GCVS&$b$&sp&$T_{eff}$&$\log g$&[Fe/H]\\

\hline

18092+2704&-&V887 Her&+20.2&F3Ib$^{1}$&6600$^{2}$&1.05$^{2}$&-0.78$^{2}$\\
19386+0155&-&V1648 Aql&-10.1&F5I$^{3}$&6800$^{4}$&1.4$^{4}$&-1.1$^{4}$\\
19500--1709&187885&V5112 Sgr&-21.0&F0I$^{3}$&8000$^{5}$&1.0$^{5}$&-0.6$^{5}$\\

\hline
\end{tabular}

\bigskip
$^{1}$ Hrivnak et al. (1988), $^{2}$ Klochkova (1995), $^{3}$
Suarez et al. (2006), $^{4}$Pereira et al. (2004), $^{5}$ Van Winckel \&\
Reyniers (2000)\\
 \end{center}
\end{table}

\section{IRAS $18095+2704$=V887 Her}

\subsection{The known properties}

This supergiant with the IR-excess was discovered by Hrivnak et
al. (1988) and classified by them as the most probable candidate
to the protoplanetary objects. They obtained also the spectral
type of the star -- F3Ib, -- and the first $BV$-estimates
including the colour excess. Van der Veen et al. (1989) give the
distance to the star of 2.4 kpc and the position above the plane
of the Galaxy at $z=800$ pc. Later Klochkova (1995) have
obtained an echelle spectrum of the star, have identified
absorption lines in the spectral range of 5050--8700~\AA, and have
estimated the model atmosphere parameters: $T_{eff}=6600$ K, $\log
g=1.0$. [Fe/H]$=-0.78$ dex, and the CNO overabundance with respect
to the solar ratio [X/Fe]$\approx 0.5$. Tamura and Takeuti (1991)
noted an inverse P Cyg-like profile of the H$\alpha$ line.
According to Klochkova  (1995), the H$\alpha$ contains the
central absorption component and two emission component,
blueshifted and redshifted ones. HST images of IRAS $18095+2704$
(Ueta et al. 2000) reveal bipolar reflective nebulosity of
$3^{\prime \prime}$ in diameter.

\subsection{Our $UBV$-observations}

The starting point of the $BV$-observations of IRAS $18095+2704$
is September 3, 1987: Hrivnak et al. (1988) have obtained
$V=10.^m43$ and $B=11.^m46$.

We have started our observations of IRAS $18095+2704$ in 1990, and
for the 19 years we have obtained more than 300 individual
estimates of the star magnitudes in the $UBV$-bands. As a comparison
star, we used Tyc 2100-384-1 for which we had undertaken
calibration procedure with the Johnson's standard stars; the
results of this calibration are: $V=11.^m85$, $B=12.^m65$,
$U=13.^m01$.

Early results have been published by us in the papers by Arkhipova
et al. (1993) and Arkhipova et al. (2000). Also due to our
efforts, the star has been inserted into the General Catalogues of
Variable Stars as V887 Her with the variability type of SRD
(Kazarovetz et al. 1993).

Fig. 1 presents light curves in the $V$-band and
colours $B-V$ and $U-B$ during the full period of our
observations, 1990--2008. Against the total linear trend of the
brightness, the star demonstrates low-amplitude quasi-periodic
oscillations. After removing the trend, from the analysis of the
2000-2008 data we have found the period of the light variations
for V887 Her, $P=109 \pm 2$ days. It is consistent with the
result by Hrivnak and Lu (2000) who have found the variability of
the radial velocity of IRAS$18095+2704$ with the period of 109
days.

\begin{figure}[!h]
 \begin{center}
  \resizebox{15 cm}{!}{\includegraphics{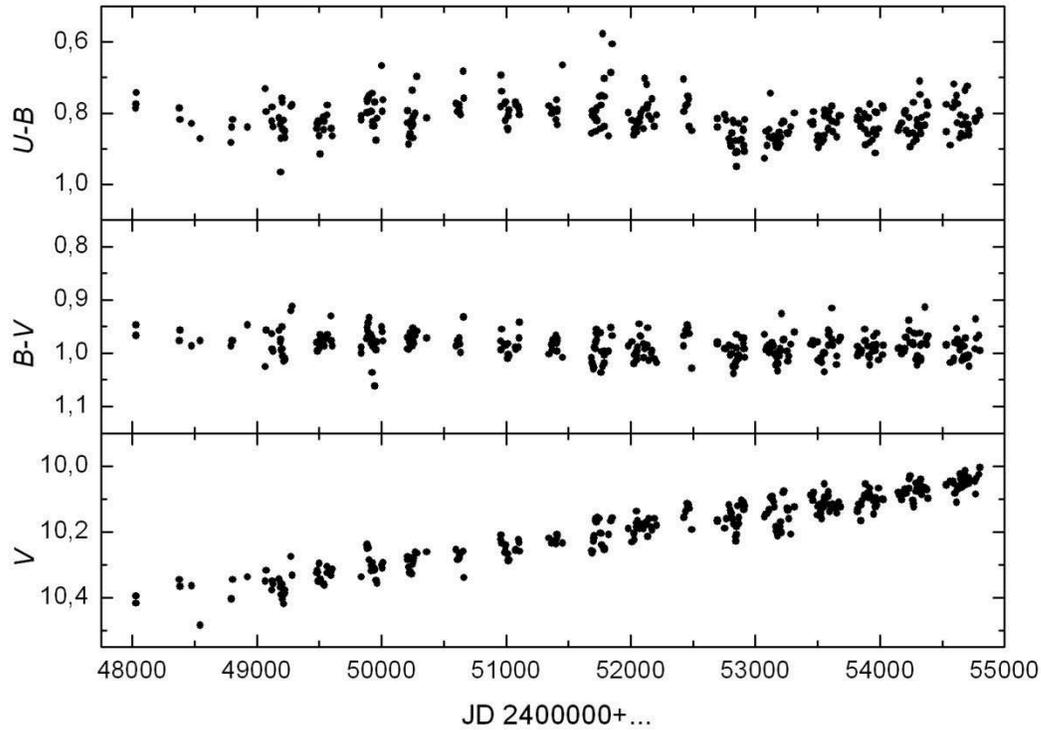}}
  \caption{The light and colours curves of V887 Her in 1990–-2008.}
  \label{v_bv_ub_1809}
 \end{center}
\end{figure}

Figure 2 shows the periodogram and phase curve for the data in
the $V$-band. The frequency corresponding to the main period of 109
days and also the annually conjugated frequencies are clearly
distinguished. The full amplitude of the wave with the period of
109 days is 0.06 mag. Though the double amplitude is always less
than 0.15 mag, the scatter of measurements at the same phase
reaches 0.1 mag. This scatter results from the fact the shape of
the light curve changes from one cycle to another; besides
quasiperiodic brightness oscillations caused by pulsations, there
is also instability due to variable stellar wind which contributes
significantly into the light variations. The variable emission
line H$\alpha$ (Tamura and Takeuti 1991, Klochkova 1995) gives
also the evidence for the continuing mass loss through the stellar
wind.

\begin{figure}[!h]
 \begin{center}
  \resizebox{15 cm}{!}{\includegraphics{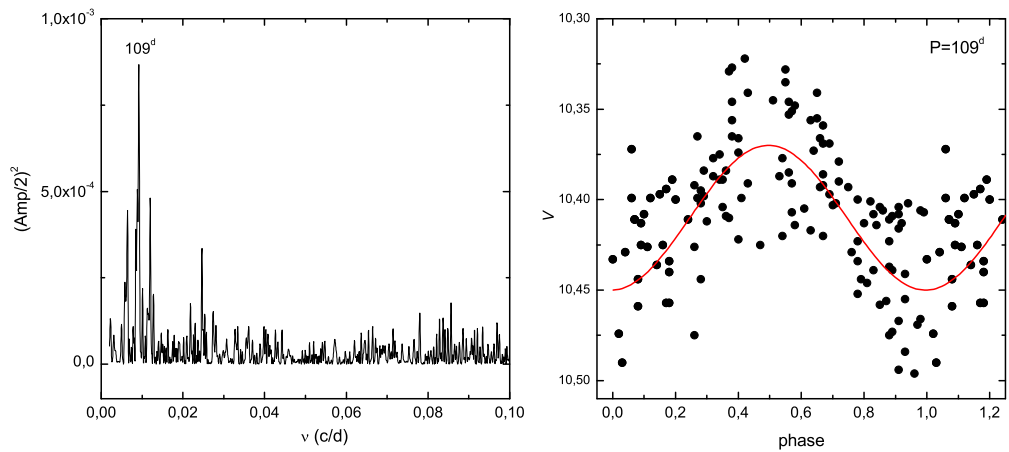}}
  \caption{The periodogram and the phased
$V$ light curve  of V887 Her.}\label{v887_phase}
 \end{center}
\end{figure}

In general, the colour variations follow the light oscillations.
During the epochs when the regular periodicity was prominent, for
example in 2004, the star became bluer while becoming brighter,
just as a typical Cepheid does. In other years, when the
instability related to the varying power of stellar wind dominated
in the total variability, correlation between the colours and the
brightness is very weak.

These low-amplitude quasiperiodic light variations of V887 Her
are observed against the systematic long-term brightness increase.
For the 19 years of our observations, the mean brightness level
has risen by 0.32 mag in the $U$-band, by 0.35 mag in the $B$-band,
and by 0.37 mag in the $V$-band, the mean colour $B-V$ being almost
constant, and the colour $U-B$ oscillating within 0.1 mag. Our
conclusion about the systematic brightness increase of V887 Her is
supported by the NIR-photometry data. According to unpublished
measurements by Yudin (1999), between 1992 and 1999 the $J$-band
brightness of V887 Her rose by 0.02 mag per year -- with the same
rate as in the $V$-band.

We can hardly explain simultaneously the constant colours and
rising total flux of V887 Her by changing parameters of the very
star, so we choose to suggest that the rising brightness of the
star is caused by changing the properties of the surrounding
envelope.

The dust-gaseous nebulosity surrounding V887 Her has a complex
structure: it looks like a disk or torus with bipolar jets (Ueta
et al. 2000). The dusty disk provides a significant reddening of
the star. By assuming spectral classification of F3I for V887 Her,
we estimate the colour excess as $E(B-V)=0.^m68$. The contribution
of the interstellar extinction into $E(B-V)$ does not exceed
$0.^m11$, according to the maps by Schlegel et al. (1998). So the
dominant part of the reddening, more than $0.^m5$, is due to the
circumstellar dust envelope. For the 19 years of observations the
mean colours of V887 Her has not become bluer so we cannot explain
the long-term brightening of the star by the growing transparency
of the envelope. We can however suggest that there are some
stratification of dust grains by sizes. Then the brightening
without blueing may be explained by dissipation of the part of the
envelope dominated by the large grains providing almost
non-selective extinction law, $R_V=A_V/E(B-V) >7$.

Some hints on the evolution of the dust envelope come from the
results of polarimetric observations. Trammell et al. (1994) found
a significant intrinsic polarization in V887 Her, $P=6.59 \pm
0.1$\%. The comparison of the measurements by Trammell et al.(1994)
in October 1991 and by Biging et al. (2006) in May 2000 shows that
while the polarization degree changed insignificantly, by only
0.8\%, the orientation of the polarizarion plane had turned
strongly, from $\theta$ =131$^{\circ}$ to $\theta$ =16$^{\circ}$
(Bieging et al. 2006). We feel that further photometric, spectral
and polarimetric observations are needed for V887 Her to clarify
the origin of the long-term light variations.

\section{IRAS $19386+0155$=V1648 Aql}

\subsection{The known properties}

The bright IR-source IRAS $19386+0155$ had been classified as a
protoplanetary object by Van der Veen et al. (1989) from the
properties of the star in the far-IR range, 12--100 $\mu$m. Those
authors were also first who observed the star in the optical and
NIR spectral range and obtained the spectral energy distribution
over the wide spectral range of 0.44--100 $\mu$m. The shape of the
spectrum and the location of the star at the two-colour diagram
($J-H$, $H-K$) implied a considerable NIR excess. The conclusion was
that the star is surrounded not only by a cold dust envelope
formed during the AGB stage but also by warm dust which was
ejected after the leaving  asymptotic-giant branch. Van der Veen
et al. estimate the rate of the mass loss as $10^{-5}\,M_{\odot}$
per year.

The full spectral analysis in the NIR and optical range has been
made by Pereira et al. (2004). They have obtained the atmospheric
parameters of the star, $T_{eff} = 6800 \pm 100$K and $\log g =1,4
\pm 0.2$, as well as the chemical abundances. The star is rather
metal-poor, [Fe/H]$=-1.1$, and the CNO-elements and also the
elements of the $\alpha$--process are underabundant. Pereira et
al. (2004) have fitted the spectral energy distribution of IRAS
$19386+0155$ by three components: a stellar photosphere, a
cold-dust spherically symmetric envelope ($T_d \approx 200 K$),
and by a hot-dust disk with the temperature of $\sim$ 1000 K.

The presence of the dust disk is often thought to be a signature
of the binarity. However up to now no traces of the second star
have been revealed, either in the optical spectrum nor in the
$\lambda$2.4--5 $\mu$m spectrum which has been obtained by the
space telescope ISO equipped with the spectrograph SWS (Pereyra et
al. 2004).

Su\'{a}rez et al. (2006) have determined a spectral class of F5I in
the spectral range of 3800--9300~\AA\ by using a low-resolution
spectrum. Gledhill (2005) have obtained direct and polarized
images of IRAS $19386+0155$ in the $J$-band. A nebulosity of
$2^{\prime \prime}$ in diameter is found which disperses the light
of the star. The polarization degree of the star light is
extremely low, $P<2$\%. Gledhill (2005) has made a preliminary
conclusion about the bipolar geometry of the nebulosity.

\subsection{Our $UBV$-observations}

The starting point of the $BV$-observations of IRAS $19386+0155$
is June 16, 1987: Van der Veen et al. (1989) have obtained
$V=11.^m21$ and $B=12.^m24$.

We began to observe IRAS $19386+0155$ in May 1990, and to the fall
of 2008 had obtained more than 240 estimates of the star
brightness in the $UBV$-bands. The comparison star was
Tyc 483-1122-1, which had $V=11.^m32$, $B=12.^m19$, and $U=12.^m40$
according to our data.

Just the first observations in 1990--1992 had revealed variability
of the light of IRAS $19386+0155$ (Arkhipova et al. 1993), which
was confirmed by the following observations (Arkhipova et al.
2000), and the star was included into the General Catalogue of
Variable Stars as V1648 Aql.

\begin{figure}[!h]
 \begin{center}
  \resizebox{15cm}{!}{\includegraphics{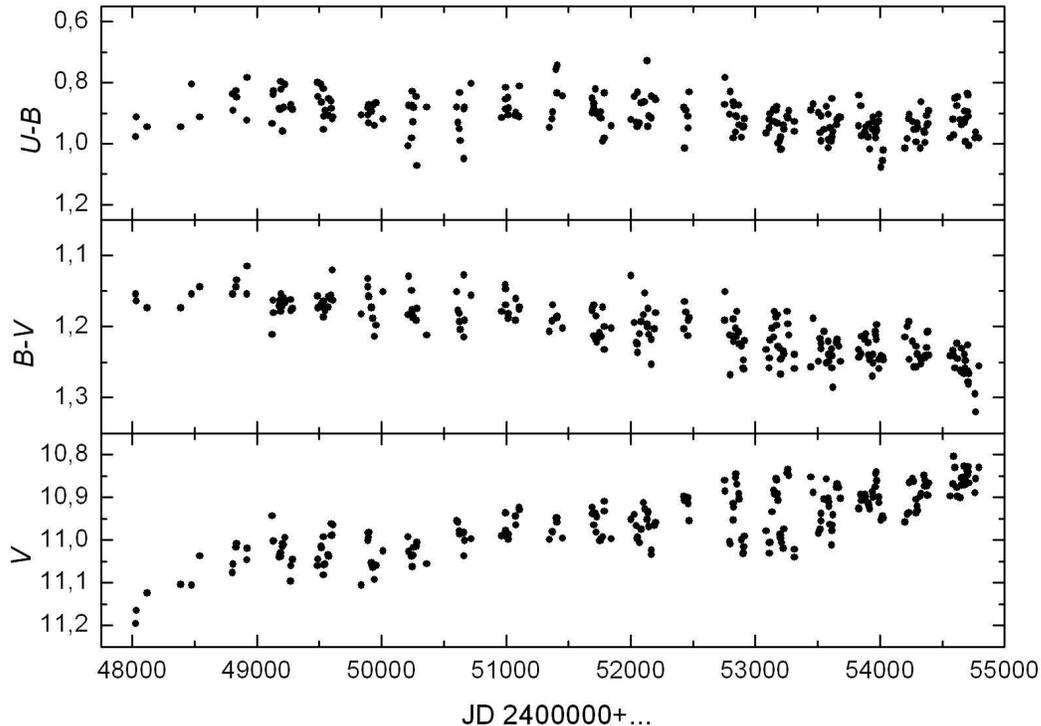}}
  \caption{The light and colours curves of V1648 Aql in 1990–-2008.}\label{vbvub1648}
 \end{center}
\end{figure}

Fig. 3 presents time variations of the $V$-brightness
and of the colours $B-V$ and $U-B$ between 1990 and 2008. The star
V1648 Aql demonstrates sine-like variations of the light with
varying amplitude; in particular, the $V$-band light curve reveals
a clear amplitude modulation.  The maximum amplitudes are $0.^m2$
in $V$, $0.^m25$ in $B$, and $0.^m4$ in $U$. In course of the
pulsations a correlation between the $V$-brightness and the $B-V$
colour is observed: while the brightness increases, the star
becomes bluer.

We searched for periodic oscillations  over our $UBV$-data of
2000--2008. Figure 4a presents the periodogram for the $V$-data and
the phase curve convolved with the period of $102^d$. The
periodogram is dominated by the peaks corresponding to $P=102^d$
and to its annually conjugated periods $P=80^d$ and $P=141^d$. We
have subtracted the wave with the period of $102^d$, and in the
residual light variations we have found periods of $98^d$ and
$106^d$. The phase curve for the more significant period of two,
$98^d$, and the periodogram for the residual light variations
are presented in Fig. 4b. Analysis of the data in the $B$ and
$U$-bands has given the same values for the periods. The frequency
ratio between the 98-days wave and the 102-days wave is 1.04. We
conclude that both periods are real ones, and the light variations
of V 1648 Aql at two close frequencies result in the amplitude
modulation which is the most prominent in the $V$- and $B$-bands.

Interestingly, our observations of 1990-1999 implied the period of
$98.8^d$ (Arkhipova et al. 2000), and Hrivnak \&\ Lu (2000) from
their observations of 1994--1995 have found $P=96^d$. Meantime the
analysis of the ASAS data for 2003--2004 which we have undertaken
reveals a period of $P=102^d \pm 1^d$. We cannot say yet if the
main period grows with time and what is the reason of simultaneous
oscillations at two close frequencies.

Besides the quasiperiodic light oscillations, V1648 Aql
experiences systematic increase of the annually averaged
brightness and the reddening in $B-V$: for the 19 years the star
becomes brighter by $\Delta V=0.^m2$ and redder by $\Delta
(B-V)=0.^m1$; the colour $U-B$ demonstrates only a weak tendency
to rise. We have estimated the colour excess for V1648 Aql as
$E(B-V)=0.^m8 - 0.^m9$, by assuming the spectral type of F3--F5I
with the normal colour of $(B-V)_0 =0.^m3 - 0.^m4$. The maximum
interstellar colour excess in this direction according to the maps
by Schlegel et al. (1998) is $E(B-V)_{IS}=0.^m41$. So a
significant part of the colour excess, $E(B-V)_{CS}\approx 0.^m4$,
is due to the circumstellar dust envelope.

When a star has a dusty disk around it, its systematic reddening
may be caused either by changing its own properties or by changing
the properties of the surrounding dust.

If the trends of the brightness and colour of V1648 Aql are due to
temperature effect during the evolution of the star along the
horizontal track at the H-R diagram, then the reddening which we
observe together with the brightening of the star in the $V$-band
taking into account its spectral type of F3-F5I must signify the
temperature decrease. According to the theoretical calculations by
Bl\"{o}cker (1995), a duration of the post-AGB stage may be 10$^2$--10$^4$
years, depending on the initial mass of a star and on the history
of its mass loss. The post-AGB evolution of the most massive stars
can be observationally traced over the man-life time intervals,
during a few dozen years. According to the star evolution theory,
photometric changes due to the evolution of a F-type
AGB-supergiant must be blueing and brightness decrease; meanwhile
we observe the $V$-band brightening and the reddening of V1648 Aql
which means that the star temperature decreases, and the star
moves to the right at the H-R diagram. Reverse evolutionary tracks
at the post-AGB stage are not excluded in the evolution theory but
they may take place only for the stars at the latest stage of
helium flash before becoming a planetary nebula central star. The
only known stars at this evolutionary stage in our Galaxy are FG
Sge, V605 Aql, and V4334 Sgr.

\begin{figure}[!h]
 \begin{center}
  \resizebox{13cm}{!}{\includegraphics{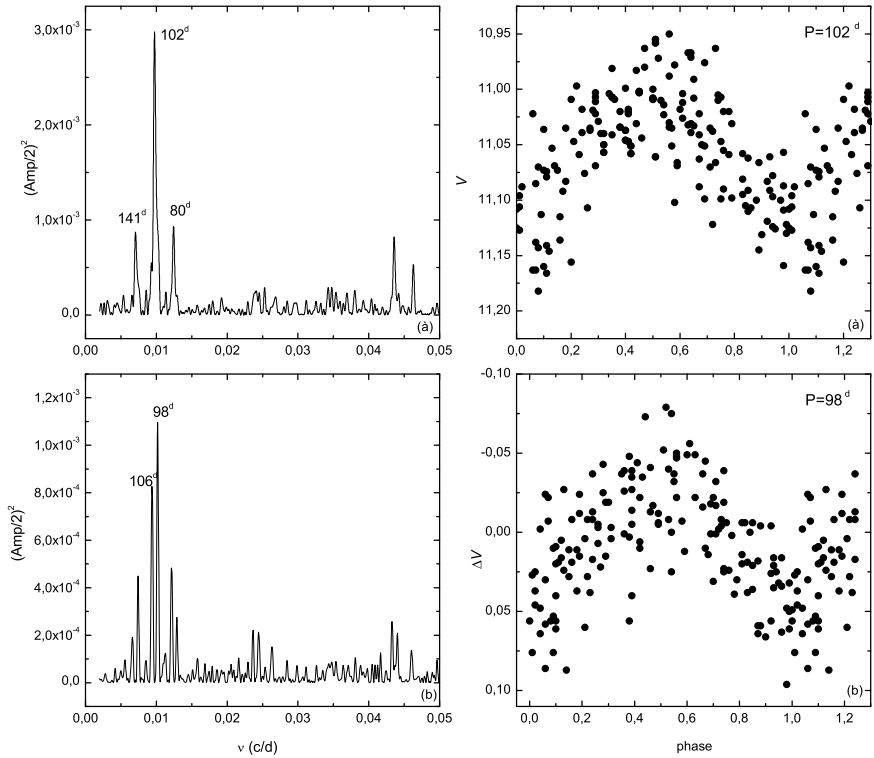}}
  \caption{The periodogram and the phased
$V$ light curve of V1648 Aql for observations in
1990-2008 (top panels -- (a)) and the periodogram and the phased
$\Delta V$ light curve for the residual data after subtraction of the wave with
the period $P$=102$^{d}$ (bottom panels -- (b)).}

\label{phase1648}
 \end{center}
\end{figure}

To check the hypothesis of possible eye-traced evolution of V1648
Aql under a constant bolometric luminosity toward the H-R diagram
area of cold stars we have estimated the temperature-induced flux
change corresponding to the normal-colour change of V1648 Aql by
$\Delta (B-V) =0.07^m$ for 15 years. The bolometric-correction
difference for a F5I-type star appears to be $\Delta BC = 0.02$
corresponding to the $\Delta (B-V)=0.07$, that results in $V$-band
brightening by $0.^m02$ -- significantly less than the observed
$V$-band brightening by $0.^m12$.

We think that the brightnening of V1648 Aql accompanying by the
reddening may be caused by an episode of intense mass loss which
has increased an opacity of the stellar atmosphere; due to this we
see now the higher and colder levels of the star. We know another
star with the similar photometric behaviour, HD 179821 (Arkhipova
et al. 2009), but the evolutionary status of HD 179821 is even
more uncertain than that of V1638 Aql: equally well it may be a
post-AGB star or a massive hypergiant. But it is evident that the
only 19 years of photometric observations of V1648 Aql are
insufficient to make certain conclusions about the nature of its
long-term light and colour trends.

\section{IRAS $19500-1709=\mbox{HD} 187885=V5112$ Sgr}

\subsection{The known properties}

The bright (according to BD, $m_{vis}=8.^m9$) star BD-17$^{\circ}$
5779=HD 187885, with the spectral class of F8 (HD) and high
latitude of $b=-21^{\circ}$, was detected by IRAS as having
FIR-light excess (Parthasarathy and Pottasch 1986). In the IRAS
catalogue the star is listed as IRAS $19500-1709$. By using the
IRAS data,  Likkel et al. (1987) identified this star as a
protoplanetary object, together with some other supergiants at
high galactic latitudes. Pottasch and Parthasarathy (1988) had
explained the IR-excess by the radiation of cold dust with the
temperature of 100 K.

Later Van Winckel et al. (1996) analysed a high-resolution spectrum
of HD 187885, determined its chemical abundance and calculated an
atmosphere model. The low metallicity ([Fe/H]$=-0.5$), the
CNO-overabundance, and the overabundance of the s-process elements
give evidence for the post-AGB evolutionary status of the star. In
this paper Van Winckel et al. considered two atmosphere models for
the star, one with $T_{eff}=8000$K, $\log g =1.0$ and another with
$T_{eff}=7700$K, $\log g =0.5$; later Van Winckel \&\ Reyniers
(2000) chose the high-temperature model of two.

A prominent peculiarity of HD 187885 is inverse P Cyg-like profile
of its spectral line H$\alpha$ detected by Tamura \&\ Takeuti
(1991).

Spectral classification of HD 187885 is somewhat uncertain and
depends on the method and criteria used. The HD catalogue gave the
spectral type of F8 for the star, however all the later
classifications were much earlier: F2-3I by Parthasarathy et al.
(1988), F2-5I by Volk and Kwok (1989), F0Ie by Su\'{a}rez et al.
(2006). The estimate based on the Geneve photometric indices gives
even earlier spectral classification: A2-3I (Van Winckel et al.
1996).

There are some indications of the binary status of the star.
Likkel et al. (1987) refering to unpublished data by Sanduleak
classified HD 187885 spectral class as a composite $A+K$.
Parthasarathy et al. (1988) noted that while the spectrum in the
optical range, $\lambda$3500--7400~\AA, corresponds to the
spectral class of F2-3, the UV-flux detected by the IUE indicates
the presence of a hotter source, with the temperature of 8500 K.

\subsection{Our $UBV$-observations}

The starting point of the $UBV$-observations of IRAS $19500-1709$
is September 4, 1987: Hrivnak et al. (1989) have published
$V=8.^m67$, $B=9.^m19$, and $U=9.^m34$.

We are monitoring HD 187885 in the $UBV$-bands starting from 1993.
As a comparison star, we used BD--17$^{\circ}$5778 for which we
had undertaken calibration procedure with the Johnson's standard
stars; the results of this calibration are: $V=9.^m97$,
$B=10.^m59$, $U=10.^m68$. For the 16 years we have obtained more
than 220 individual photometric estimates. We have revealed the
star variability by analysing the data of 1993-1999 (Arkhipova et
al. 2000), and HD 187885 has been included into the General
Catalogue of Variable Stars as V5112 Sgr (Kazarovets et al. 2003).
Our observations after 1999 have allowed us to study in detail the
character of low-amplitude quasiperiodic variability of the star
and also to confirm long-term trends of its brightness.

\begin{figure}[!h]
 \begin{center}
  \resizebox{15cm}{!}{\includegraphics{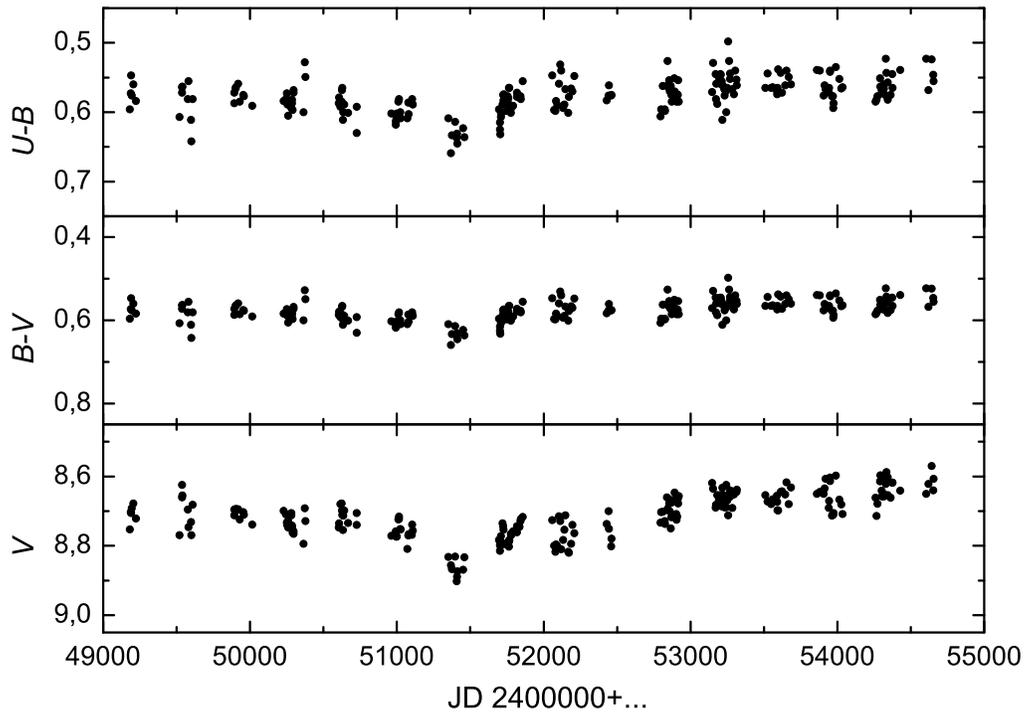}}
  \caption{The light and colours curves of V5112 Sgr in 1993-2008.}\label{v_bv_ub_1630}
 \end{center}
\end{figure}

\begin{figure}[!h]
 \begin{center}
  \resizebox{10cm}{!}{\includegraphics{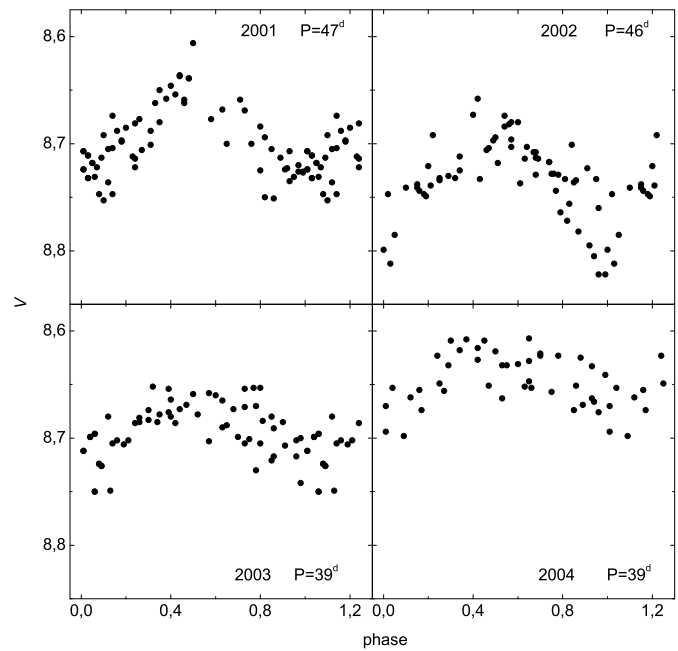}}
  \caption{The ASAS phase light curves of V5112 Sgr.}\label{phase_1630}
 \end{center}
\end{figure}

Fig. 5 presents the light curve and colours
variations traced by the star observations in 1993--2008. The
variability within a single year can be characterized as
low-amplitude ($\Delta V <0.^m15$) quasiperiodic oscillations with
varying shape of the light curve. We searched for a variability
period over every observational year separately, starting from
2001. To increase the statistics, we have also involved the data
of the ASAS survey for 2001--2004. Table 2 presents a list of the
periods obtained from the observational data of 2001-2008. Two
periods of the V5112 Sgr light variations are visible in different
epochs: $38.7 \pm 0.4$ and $47.5 \pm 0.5$ days. The ratio of these
two periods is close to 0.8. Figure 6 shows the phase curves from
the ASAS data of 2001--2004.

\begin{table}
\begin{center}
\caption{The periods of V5112 Sgr}
\bigskip
\begin{tabular}{|c|c|c|}
\hline
Years&$P$, d (ASAS)&$P$, d (our\\
&&data)\\
\hline
2001&47.6, 39.0&47.0\\
2002&46.1&-\\
2003&39.1&38.6\\
2004&39.1&-\\
2005&-&38.5\\
2006-2008&-&38.0\\

\hline
\end{tabular}
\end{center}
\end{table}

The light oscillations of V5112 Sgr demonstrate a strong
correlation between the brightness and the colour: the
star becomes bluer while it brightens.

We have estimated the colour excess of V5112 Sgr as $E(B-V)=0.^m4$
under the assumption that the spectral class of the star is F0I.
The maximum interstellar extinction in this direction is not more
than $E(B-V)_{IS}=0.2$ according to the maps by Schlegel et al.
(1998). It means that about the half of all reddening is provided
by the circumstellar dust envelope.

The secular light trend of V5112 Sgr can be characterized as
follows. Between 1993 and 1997 the mean brightness of the star
stayed constant, in 1998 a decrease started, and in 1999 the star
had reached the minimum which persisted over the whole annual
visibility season. The depth of the minimum was $0.^m15$ in the
$V$-band. In 2000 the star had returned to the level of 1998, and
starting from this year the brightness of the star increased
monotonically. The brightening was accompanied by the blueing: the
star had the reddest colour during the minimum of 1999.

Since there are some suspicions about the binarity of V5112 Sgr,
we can suggest that the light minimum of 1999 has been in fact an
occultation. To check this hypothesis, we need more time because
during the 16 years of observations there are only one such event
which may be treated as the occultation.

\section{Discussion}

What may be the reasons and manifestations of the variability of
the stars which we study, taking into account their evolutionary
status?

Theoretical calculations predict pulsational instability of the
evolved stars of intermediate masses over a wide range of
temperature (Gautschi 1993, Zalewski 1993). The F-supergiants with
the IR-excesses of our sample demonstrate low-amplitude ($\Delta V
< 0.^m2$) quasiperiodic variability with a strong correlation
between the brightness and the colour: while brightening, the
stars become bluer. The stars V887 Her and V1648 Aql have similar
effective temperatures, 6600 K and 6800 K correspondingly, and consequently
close periods of the light oscillations, 109 days and 102 days,
which is a signature of practically the same luminosities of these
stars. The star V5112 Sgr has a higher effective temperature,
$T_{eff}=8000$K, so the timescale of its light oscillations is
much shorter -- 39 and 47 days, and the periodic component is less
prominent.

After major mass loss events at the previous evolutionary stages,
at the post-AGB stage the stars continue to lose their mass via
unstable stellar wind. The manifestations of this stellar wind are
the presence of the variable emission line H$\alpha$ in the
spectra of the most post-AGB supergiants and photometric peculiar
behaviour -- chaotic low-amplitude light variability superposing
the regular pulsations. Among the stars considered in this paper,
V887 Her suffers the variability caused by the unstable stellar
wind most of all.

Post-AGB supergiants are surrounded by dust envelopes which may
contribute significantly into their total light reddening. All
three stars in our work demonstrate the dominant contribution of
the circumstellar envelopes into their reddening. The long-term
expansion of the envelopes and the subsequent drop of their
optical depth result in the brightening of the stars. The
long-term linear light trend of V887 Her under the constant
colours can be explained by the optical depth drop in the part of
the circumstellar dust envelope, just that part which consists of
rather large grains providing practically non-selective extinction
law. Therefore we have to suggest that the dust envelope of V887
Her contains a mixture of grains of various sizes, and the larger
grains destroy with time.

One of our stars, V1648 Agl, has shown systematic changes of the
light, colours, and perhaps the pulsational periods for the 19
years of monitoring. Our attempt to explain the long-term
variability of the star by the temperature decrease  due to star
evolution along the constant-luminosity part of the post-AGB
evolutionary track at the H-R diagram has been unsuccessful. We
have concluded that the most probable cause of this variability
may be intense mass loss which increases opacity of the star
atmosphere and move the photosphere up to the higher and colder
level to produce the redder and fainter continuum.

We know that a significant part of candidates to protoplanetary
objects are binary stars. For example, in the catalogue by Szczerba
et al. (2007) of the most probable post-AGB objects the stars
mentioned as binaries contribute 14\%\ of all stars (45 of 326).
One of these suspected binaries is V5112 Sgr studied by us in this
work; it has experienced an annual drop of the light which may be
interpreted as an occultation in the binary system.

\section{Conclusions}

We can summarize our prolonged photometric observations of three
candidates to protoplanetary nebulae as follows.

The F3-supergiant IRAS $18095+2704=V887$ Her demonstrates
quasiperiodic oscillations with the amplitudes up to $\Delta
V=0.^m15$, $\Delta B=0.^m20$, $\Delta U=0.^m25$. The pulsational
light oscillations with the period of $109\pm 2$ days are
superposed with the chaotic light variations due to the stellar
wind variability. The long-term linear trend of the light -- the
brightening of the star with the rate of $0.^m02$ per year under
the constant mean colours -- may be a result of the increasing
transparency (or dissolution) of the dust envelope consisting of
rather large grains providing non-selective extinction.

The 2000--2008 light curve of the F5-supergiant IRAS
$19386+0155=V1648$ Aql which has significant NIR and FIR light
excesses can be fitted by a sum of two waves with the periods of
$102^d$ and $98^d$ and full amplitudes of $0.^m11$ and $0.^m07$,
respectively. The oscillations at two close frequencies results in
the amplitude modulation. Besides these short-term variations, for
the 19 years of observations V1648 Aql becomes persistently
brighter by $0.^m2$ in the $V$-band, the colour $B-V$ has
increased by $0.^m1$, and the colour $U-B$ has also shown a weak
tendency to rise.

The hottest of our stars, IRAS $19500-1709=V5112$ Sgr, has
demonstrated the shortest period of the pulsations and the unusual
trend of annually averaged brightness. The star experiences
semiregular light variations with the amplitudes up to $0.^m15$ in
the $V$-band and with the cycle durations of 39 and 47 days. The
ratio of two quasiperiods is 0.8. During the 16 years we traced
also long-term variations of the light and colours of the star. In
1999 the star has reached a minimum of its brightness. If it is an
occultation in a binary system, the orbital period has to exceed
6000 days.

\section*{Acknowledgements}

We thank V.P. Goranskij and N.N. Samus for their interest to our
work and useful comments.

The work is partly supported by the grant of the President of
Russian Federation for the state support of leading scientific
schools NSch-433.2008.2.102.

 \bigskip

 \bigskip
\section*{References}

Arkhipova V. P. , Ikonnikova N. P., and  Noskova R. I., Pis'ma
Astron. Zh. {\bf 19}, 436 (1993) [Astron. Lett. {\bf 19}, 169
(1993)].

Arkhipova V. P.,  Ikonnikova N. P.,  Noskova R. I. and  Sokol G.V.,
Pis'ma Astron. Zh. {\bf 26}, 705 (2000) [Astron. Lett. {\bf 26},
609 (2000)].

Arkhipova V. P.,  Esipov V. F.,  Ikonnikova N. P.,
Komissarova G. V.,  Tatarnikov A. M., and  Yudin B. F., Pis'ma Astron. Zh.
(2009) {\bf 35}, 846 [Astron. Lett. {\bf 35}, 764 (2009)].

Bieging J.H.,  Schmidt G.D.,  Smith P.S. and  Oppenheimer B.D.
Astrophys. J., {\bf 639}, 1053 (2006).

Bl\"{o}cker T., Astron. Astrophys., {\bf 299}, 755 (1995).

Gautschy A., Monthly Notice Royal Astron. S. {\bf 265}, 340
(1993).

Gledhill T.M., Mon. Not. R. Astron. Soc., {\bf 356}, 883 (2005).

Hrivnak B.J.,  Kwok S.,   Volk K.M., Astrophys. J., {\bf 331}, 832
(1988).

Hrivnak B.J.,  Kwok S.,   Volk K.M., Astrophys. J., {\bf 346}, 265
(1989).

Hrivnak B.J. \&\ LU W.X. , 177th Symposium of the IAU, held in
Antalya, Turkey, 27-31 May, 1996. Ed. by Robert F. Wing, "The
carbon star phenomenon", {\bf 177}, 293 (2000).

Kazarovets E.V.,  Samus N.N., Goranskij V.P., IBVS, No 3840, 1
(1993).

Kazarovets E.V.,  Kireeva N.N.,  Samus N.N. and  Durlevich O.V.,
IBVS, No 5422, 1 (2003).

Klochkova V.G., Mon. Not. R. Astron. Soc., {\bf 272}, 710 (1995).

Likkel L.,  Omont A.,  Morris M.  and  Forveille T., Astron.
Astrophys., {\bf 173}, L11-L14 (1987).

Parthasarathy M. \&\ Pottasch S.R., Astron. Astrophys., {\bf 154},
L16 (1986).

Parthasarathy M.,  Pottasch S.R. and Wamsteker W., Astron.
Astrophys., {\bf 203}, 117 (1988).

Pereira C.B.,  Lorenz-Martins S. and  Machado M., Astron.
Astrophys., {\bf 422}, 637 (2004).

Pojmanski G., Acta Astronomica,  {\bf 52}, 397 (2002)

Pottasch S.R. \&\ Parthasarathy M., Astron. Astrophys., {\bf 192},
182 (1988).

Schlegel D. J.,  Finkbeiner D. P.,  Davis M., Astrophys. J., {\bf
500}, 525 (1998).

Su\'{a}rez O.,  Garc\'{i}a-Lario P.,  Manchado A.,  Manteiga M.,  Ulla A.,
S. R. Pottasch, Astron. and Astrophys. {\bf 458}, 173 (2006).

Szczerba R.,  Siodmiak N.,  Stasinska G. and  Borkowski J., Astron.
Astrophys., {\bf 469}, 799 (2007).

Tamura S.  \&\ Takeuti M., IBVS, No 3561 (1991).

Trammell S.R. , Dinerstein H.L. and Goodrich R.W., Astron. J., {\bf
108}, 984 (1994).

Ueta T.,  Meixner M. and Bobrowsky M., Astrophys. J., {\bf 528},
861 (2000).

Van der Veen W.E.C.J., Habing H.J. and Geballe T.R., Astron.
Astrophys., {\bf 226}, 108 (1989).

Volk K.M. \&\ Kwok S., Astrophys. J., {\bf 342}, 345 (1989).

Van Winckel H., Waelkens  C. and Waters L.B.F.M., Astron.
Astrophys., {\bf 306}, L37 (1996).

Van Winckel H. \&\ Reyniers M., Astron. Astrophys., {\bf 354}, 135
(2000).

Yudin B.F., private communication (1999).

Zalewski J., Acta Astronomica, {\bf 43}, 431 (1993).

\end{document}